\documentclass[nofootinbib, twocolumn, prd]{revtex4-1}
\usepackage{hyperref}
\usepackage[utf8]{inputenc}
\usepackage{amsmath}
\usepackage{amssymb}
\usepackage{courier}
\usepackage{listings}
\usepackage{color}
\usepackage{tabularx,colortbl}
\usepackage{graphicx}
\usepackage{fancyref}
\usepackage{tensor}
\usepackage{appendix}

\begin{document}

\title{Generalizing weak gravity conjecture}
\author{Rance Solomon}
\email{rancesol (at) buffalo.edu}
\author{Dejan Stojkovic}
\email{ds77 (at) buffalo.edu}
\affiliation{HEPCOS, Department  of  Physics,  SUNY  at  Buffalo,  Buffalo,  NY  14260-1500, USA}

\begin{abstract}
        The weak gravity conjecture implies the necessary existence of particles with charge-to-mass ratio $q/m \geq 1$ so that the extremal charged black hole can completely evaporate without leaving a dangerous stable extremal remnant while simultaneously not revealing a naked singularity along the way.
        In other words, this inequality ensures that the charge is emitted faster than the mass of a black hole, which is in turn coincidentally consistent with the fact that gravitational interaction for such parties is weaker than electromagnetic.
        To extend this argument to non-extremal black holes, we solve the problem of a charged shell of mass and charge ($m,q$) from a black hole with mass ($M,Q$). We find a more general condition $q/m \geq Q/M$, which obviously reduces to the weak gravity conjecture in the extremal limit, however it relaxes the condition for complete evaporation of non-extremal black holes. This condition also allows us to directly relate the particle content of the theory with the spectrum of black hole states.  
\end{abstract}

\maketitle

\section{Introduction}

The Reissner-Nordstr{\"o}m (RN) solution describes a spacetime geometry sourced by a central mass $M$ and a $U(1)$ gauge charge, $Q$ (e.g. electromagnetism). When the mass and charge distribution are both contained below the horizon radius it is understood that a black hole (BH) is formed.
It is an interesting feature of the RN solution that the geometry houses two event horizons for well behaved values of $Q$ and $M$ located at radii
\begin{equation}\label{eq:RN_radii}
    R_{\pm}=M\pm\sqrt{M^{2}-Q^{2}}.
\end{equation}
Here, and in the remainder of this article, we have used $m_{Pl}=1$ such that $Q$ and $M$ are of the same dimension.
As we can see, the Schwarzschild solution is obtained by taking $Q \to 0$.
On the other hand, an extremal BH occurs at the bound $Q = M$. Beyond the extremal bound, $Q>M$, the RN solution no longer houses a horizon and possibly violates the cosmic censorship conjecture (CCC).

It is evident then, if the CCC is not to be violated, that the BH evaporation process (whether through Hawking evaporation, Schwinger effect, or some other mechanism) should not be able to continue past the extremal bound.

We could consider a mechanism that stops the evaporation processes at the extremal bound (much like how Hawking evaporation already naturally does) and allow the extremal BH to form a remnant, stable against further decay.
But stable BH remnants are not preferred for at least two reasons:
\begin{enumerate}
    \item The base $\Lambda$CDM model has that the early universe, post inflation but pre-BBN, was radiation dominated so that the equation of state for the universe was close to $w=1/3$ at BBN. Also at this time, primordial BHs (PBHs) were able to form in abundance with an effective equation of state $w=0$ \cite{Carr:1975qj}. So to keep in agreement with BBN predictions, the majority of PBHs must have then evaporated away into some massless degrees of freedom. But for PBHs which have taken on some net charge, if the evaporation process only goes down to the extremal bound, stable relics form and keep the equation of state away from $w=1/3$, or they can even overclose the universe at early times (see \textit{e.g.} \cite{Dai:2009hx}).
    
    \item The covariant entropy bound (CEB) \cite{Bousso:1999xy} places a fundamental limit on the degrees of freedom allowed in a theory. Violating the bound could allow for infinite entropy in some finite volume. Since each stable object able to be formed in a theory provides a degree of freedom, the CEB says that the number of stable objects a theory should be able to form must be finite. But extremal BHs could have any charge or mass satisfying $Q/M=1$, naively allowing for an infinite spectrum of degrees of freedom and violating the CEB.\footnote{The concern here is not entirely unanimous in the literature since the discretization of charge could lead to a large but finite number of stable states within a mass range.}
\end{enumerate}
This is where the original BH arguments of the weak gravity conjecture (WGC) come in (see \cite{ArkaniHamed:2006dz}). To keep from forming dangerous stable BH remnants while simultaneously not violating the CCC, the WGC proposes that extremal BHs must decay through some mechanism in such a way that they satisfy
\begin{equation}\label{eq:inequality}
    \dfrac{Q}{M} \geq \dfrac{Q'}{M'}
\end{equation}
where ($M,Q$) and ($M',Q'$) are the mass and charge of the BH before and after the decay, respectively. 
In order to reduce the BH charge, the decay process must produce particles in the $U(1)$ particle spectrum. Thus, the arguments of the WGC say, for an extremal BH evaporation to occur, that there must be at least one particle in the spectrum with mass and charge ($m,q$) that when far removed has a net repulsive force from an extremal BH. That is to say
\begin{equation} \label{eq:original_WGC}
    \dfrac{q}{m} \geq 1,
\end{equation}
where we are using $q$ and $Q$ as the magnitudes of the charges but assume them to be of like charge. The WGC has been very useful in labeling unrealistic theories of quantum gravity by checking if they allow for such a state

Another simple argument for the WGC was discussed by Cheung and Remmen in \cite{Cheung:2014vva} where they consider the BH with ($M,Q$) which completely decays away into a final state of $n = Q/q$ particles with ($m,q$). Therefore, from energy conservation we require $M \geq nm$, or, defining $z \equiv q/m$ and $Z \equiv Q/M$,
\begin{equation} \label{eq:cheung_result}
    z \geq Z.
\end{equation}
Cheung and Remmen further argue that it is not sufficient for just one particle in the spectrum to satisfy this condition but that some weighted average of the available species must.
We take a similar stance here. But while \cite{Cheung:2014vva} concerns the charge-to-mass ratio of the entire final state, we break up the final state into thin charged shells and show that each shell, when emitted, must satisfy the bound as well.

It is worth noting that, including higher order derivative terms in the RN solution, the extremal bound could become a mass dependent function, pulling away from $Q=M$ at small masses (see \cite{Kats:2006xp}). It still remains to be proven the sign of these corrections which could either allow for the extremal bound to lie somewhere in the $Q<M$ or $Q>M$ space. Some physical arguments have been made (\cite{Hamada:2018dde, Goon:2019faz, Cheung:2019cwi, Bellazzini:2019xts, Mirbabayi:2019iae}) calling for the extremal bound to be pushed towards the $Q>M$ space which would conveniently allow for over-charged BHs themselves to satisfy the WGC. For convenience we will ignore these corrections.


\section{Shell Emission}\label{sec:BH_Decay}

Consider again a decay taking us from a BH with $Q/M \leq 1$ to a BH with $Q'/M' \leq 1$. We take this to be done by emitting a charged, thin, mass shell of rest mass $m$ and charge $q$.
We could imagine the mass shell to be the dominant s-wave of a scalar field (see \cite{Page:1976df}), or as a locus of $U(1)$ charged particles -- since the WGC is a general result it should bind these decay modes as well. In the case of the locus of particles, assuming the shell to be a uniform distribution of identical particles, the charge-to-mass ratio of the shell will be the same as the individual particles making up the shell.

\begin{figure}[b]
    \centering
    \includegraphics[width=8.6cm]{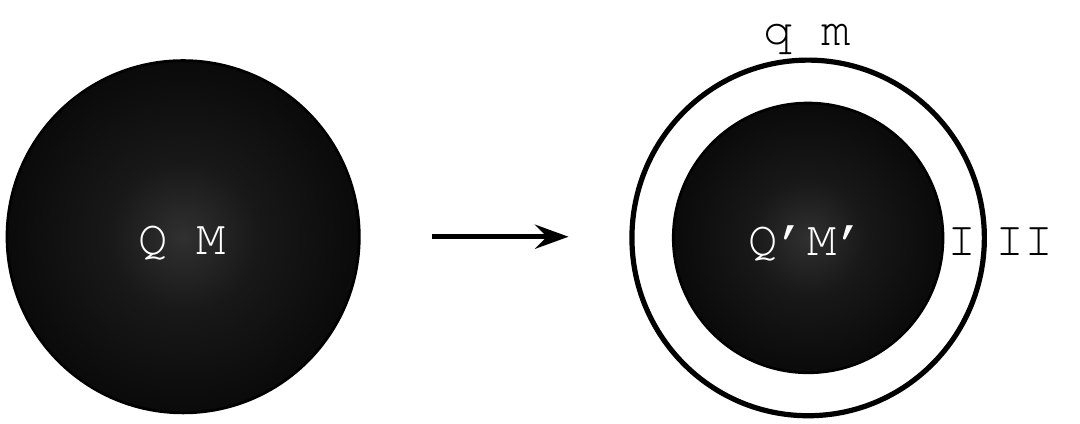}
    \caption{A BH decays into another through the emission of an outward moving thin, charged mass shell. The shell divides the space into two regions, interior (I) and exterior (II).}
    \label{fig:mass_shell}
\end{figure}

The charge of the BH before and after the decay should be related by
\begin{equation} \label{eq:charge_relation}
    Q = Q' + q.
\end{equation}
For the mass relation, we refer to Fig. \ref{fig:mass_shell}. The presence of the mass shell outside $M'$ causes a discontinuity in the extrinsic curvature tensor, $K_{ij}$, between regions I and II.
The discontinuity at the shell, denoted by the square brackets, can be found to be (see \cite{Lightman:1975} or \cite{Poisson:2009pwt})
\begin{equation}\label{eq:K1}
    [\tensor{K}{^j_i}] = 8\pi\sigma\Big(u^{j}u_{i} + \dfrac{1}{2}\tensor{\delta}{^j_i}\Big)
\end{equation}
where $\sigma$ is the mass density of the shell such that $4\pi R^{2}\sigma = m$, the rest mass of the shell, and $u$ is the 4-velocity of the shell. Considering only radial motion of the shell we can find that
\begin{equation} \label{eq:Kthetatheta}
    [K_{\theta\theta}] = 4\pi g_{\theta\theta}\sigma = 4\pi R^{2}\sigma = m.
\end{equation}
We have assumed a dust equation of state for the shell in (\ref{eq:K1}) so we should not naively take the limit $m \to 0$. Furthermore, we note that $q$ contributes to the discontinuity (\ref{eq:Kthetatheta}) only through its energy content which can be wrapped up in $m$ (see \cite{Heusler:1990in}).

We could also evaluate the discontinuity by taking the difference of the extrinsic curvature tensor inside and outside the shell,
\begin{equation*}
    [\tensor{K}{^j_i}] = \tensor{K}{^j_i}^{(II)} - \tensor{K}{^j_i}^{(I)}.
\end{equation*}
For this we consider the metric in region I to be the Reissner-Nordstr{\"o}m solution sourced by $M'$ and $Q'$,
\begin{equation*}
    ds^{2}_{I} = -f_{I}(r)dt_{I}^{2} + f_{I}^{-1}(r)dr^{2} + r^{2}d\Omega^{2}.
\end{equation*}
Likewise, region II will have a metric
\begin{equation*}
    ds^{2}_{II} = -f_{II}(r)dt_{II}^{2} + f_{II}^{-1}(r)dr^{2} + r^{2}d\Omega^{2}
\end{equation*}
where we have introduced the functions
\begin{align*}
    &f_{I}(r) \equiv \Big(1 - \dfrac{2M'}{r} + \dfrac{Q'^{2}}{r^{2}}\Big), \\
    &f_{II}(r) \equiv \Big(1 - \dfrac{2M}{r} + \dfrac{Q^{2}}{r^{2}}\Big)
\end{align*}
for brevity.
There remains a difference in time coordinates between the two regions as is necessary for the two regions to match up (see for instance \cite{Dai:2016sls}). Calculating the extrinsic curvature tensor in both regions we obtain
\begin{equation}\label{eq:discontinuity}
    [\tensor{K}{_\theta_\theta}] = -R\Big(\sqrt{f_{II}(R)+v^{2}} - \sqrt{f_{I}(R)+v^{2}}\Big) = m.
\end{equation}
where $u^{r}=v$ is the radial speed in the rest frame of the shell and $R$ is the radial size of the mass shell centered on $M'$.
Solving (\ref{eq:discontinuity}) for $M'$ we can get
\begin{equation}\label{eq:M'}
    M' = M + \dfrac{q^{2}-m^{2}-2qQ}{2R} - m\sqrt{f_{II}(R)+v^{2}}.
\end{equation}
Notice that $M$, $M'$, and $m$ are the constant ADM masses. So if we let $R \rightarrow \infty$ then we find
\begin{equation} \label{eq:mass_at_inf}
    M' = M - m\sqrt{1+v_{\infty}^{2}} ,
\end{equation}
which one might have guessed  allowing for the possibility that $m$ still has some kinetic energy far removed from the BH. Combining (\ref{eq:inequality}), (\ref{eq:charge_relation}), and (\ref{eq:mass_at_inf}) with $v_{\infty}=0$ we immediately get that
\begin{equation} \label{eq:result_at_inf}
    q/m \geq Q/M
\end{equation}
as we would expect for far removed objects. The relation in Eq.~(\ref{eq:result_at_inf}) is more general than WGC since it is valid for a general charged BH, not only extremal. This obviously reduces to the WGC, $q/m \geq 1$, in the extremal limit. But since $Q < M$ for non-extremal BHs, the condition for complete evaporation of non-extremal BHs is relaxed in comparison to WGC. In particular, particles with $q/m \leq 1$ can also reduce the $Q/M$ ratio of a BH. 

\begin{figure}
    \centering
    \includegraphics[width = 8.6cm]{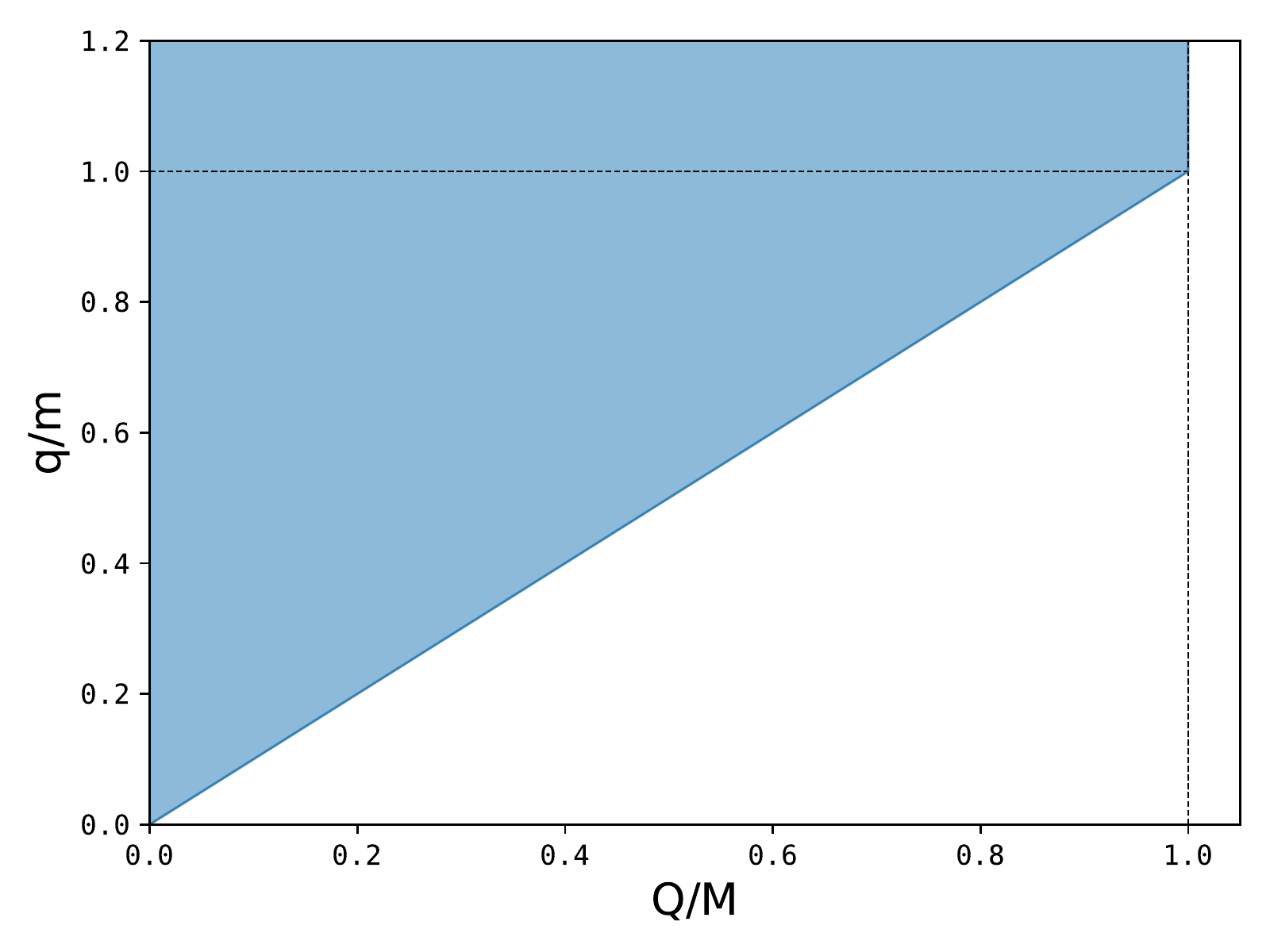}
    \caption{The shaded region marks those charge-to-mass ratios of a shell that would allow for a BH of ($M,Q$) to evaporate away from extremality. Dashed lines mark the $q=m$ and $Q=M$ bounds.}
    \label{fig:betaVepsilon}
\end{figure}

As an important consequence, the condition in Eq.~(\ref{eq:result_at_inf}) also allows us to directly relate the particle content of the theory with the spectrum of BH states.  
We show this in Fig.~\ref{fig:betaVepsilon}.

Admittedly, going from (\ref{eq:M'}) to (\ref{eq:mass_at_inf}) we assumed the shell to be able to escape from the BH which may not be possible for general values of parameters. To be more precise we can require
\begin{equation*}
    M + \dfrac{q^{2}-m^{2}-2qQ}{2R} - m\sqrt{f_{II}(R)+v^{2}} \leq M - m
\end{equation*}
which is equivalent to $M' \leq M - m$. If we consider the most extreme case where the shell originates from the horizon of the original BH, $R=M+\sqrt{M^{2}-Q^{2}}$, then solving for $v$ gives
\begin{equation} \label{eq:v}
    v \geq 1 + (z^{2}-1)\dfrac{m}{2R} - z\dfrac{Q}{R}
\end{equation}
where $z$ is defined in the same way as in (\ref{eq:cheung_result}). Thus, for the shell to escape to infinity we would require
\begin{equation*}
    2z Q \geq (z^{2} - 1)m
\end{equation*}
or equivalently
\begin{equation*}
    2qQ + m^{2} \geq q^{2}
\end{equation*}
which is trivially satisfied since $q \leq Q$. That is to say, charged shells are generally able to escape the near-horizon region of a BH, and as long as the shell's charge-to-mass ratio is at least that of the original BH's then it will allow the BH to recede from extremality.


\section{Conclusion}

In principle, if we require that BHs are allowed to evaporate completely without crossing the extremal bound we can connect the charge-to-mass ratio of the Reissner-Nordstr{\"o}m BH and the particle content in a theory.
Formerly, the WGC and its extensions have shown that at least one far removed particle -- or a combination of far removed particles -- available in a $U(1)$ spectrum must satisfy $q/m \geq 1$.
Here we have used the outflow of a charged thin shell to show that the WGC continues to hold when the decay product is far removed from the near horizon geometry and thereby we have tied the WGC explicitly to BHs in a similar sense to the BH thermodynamic arguments posed in \cite{Cheung:2018cwt} and others. This agreement has come at the requirement that the shell have at least the necessary escape velocity from the BH which in the extremal limit is found to be zero for a shell with $q/m \geq 1$ (see equation (\ref{eq:v})). 

We also wish to note that our main result,
\begin{equation}\label{mr}
    \dfrac{q}{m} \geq \dfrac{Q}{M},
\end{equation}
directly connects a general charged BH and particle spectra for a particular $U(1)$. One could perhaps conjectured or guessed this very relationship, however here we obtained it by solving an exact general relativistic problem. As an interesting consequence of Eq.~(\ref{mr}), particles with $q/m \leq 1$ can also reduce $Q/M$ ratio of a BH with $Q<M$. 

In addition, in some cases, the relation can apply to the creation of BHs in addition to their evaporation. For example, we can consider a $U(1)$ with only one particle species. If the species had $q/m < 1$ it would apparently violate the original WGC, but not our relation Eq.~(\ref{mr}). In such a theory, BHs would stay safely away from the extremal bound since collapse and/or accretion of particles with $q/m<1$ could produce only BHs with strictly $Q<M$. As a richer particle spectrum is considered a wider variety of BHs can be produced.

Furthermore, while we are in agreement with previous results, knowing that a shell of particles emitted from an extremal BH must also satisfy $q/m \geq 1$ could allow us to use  the production rates of particles to further constrain the necessary charge-to-mass ratio needed in a theory. For instance, if two charged species ($i={1,2}$) exist in a $U(1)$ then the shell will consist of some percentage of $i=1$ particles and $i=2$ particles with their contribution depending on their production rate. So if species $1$ has $q_{1}/m_{1} \ll 1$ but a very high production rate at the extremal bound compared to species $2$, that is to say $\langle N_{1} \rangle \gg \langle N_{2} \rangle$, then species $2$ would not only need $q_{2}/m_{2} \geq 1$ but instead $q_{2}/m_{2} \gg 1$. This could in turn further narrow down the allowed particle content in the theory. This idea was inspired by the work done in \cite{Hiscock:1990ex} for the electron-positron pair production through the Schwinger effect.

\section*{Acknowledgements}
The authors would like to thank D. Dai and W. Kinney for help in crucial parts of the project, and  Lubos Motl for very useful comments. D.S. is partially supported by the US National Science Foundation, under Grant No. PHY-1820738.

\end{document}